\documentclass[sigconf,screen]{acmart}
\usepackage{graphicx} 
\usepackage{booktabs}
\usepackage{hyperref}
\usepackage{enumitem}
\usepackage{amsmath}
\usepackage{makecell}
\usepackage{wrapfig}
\usepackage[most]{tcolorbox}
\usepackage[edges]{forest}
\usepackage{xcolor}
\usepackage{microtype}

\let\oldcite\cite
\renewcommand{\cite}{\,\oldcite}

\newcommand{\miniparagraph}[1]{\textit{#1}.\ }

\newcommand{\perceived}[0]{\texttt{perc}}
\newcommand{\objective}[0]{\texttt{obj}}
\newcommand{\perceivedobjective}[0]{\texttt{perc\&obj}}

\newlist{questions}{enumerate}{2}
\setlist[questions,1]{label=\textbf{RQ$_\text{\arabic*.}$},ref=RQ$_\text{\arabic*}$, leftmargin=*}
\setlist[questions,2]{label=\textbf{1.\arabic*},ref=\thequestionsi(\arabic*), leftmargin=0cm}
\newcommand{\ques}[1]{\texorpdfstring{RQ$_{#1}$}{RQ #1}}

\title{Fragmented Markers, Mixed Results: A Systematic Review of AI Coding Assistants and Developer Productivity}

\author{Annemarie Wittig}
\correspondingauthor
\orcid{0000-0002-7051-2242}
\affiliation{%
  \institution{Leipzig University}
  \city{Leipzig}
  \country{Germany}
}
\email{awittig@cs.uni-leipzig.de}

\author{Alina Mailach}
\orcid{0000-0001-6204-2095}
\affiliation{%
  \institution{Leipzig University}
  \city{Leipzig}
  \country{Germany}
}
\email{alina.mailach@cs.uni-leipzig.de}

\author{Norbert Siegmund}
\orcid{0000-0001-7741-7777}
\affiliation{%
  \institution{Leipzig University}
  \city{Leipzig}
  \country{Germany}
}
\email{norbert.siegmund@cs.uni-leipzig.de}



\copyrightyear{2026}
\acmYear{2026}
\setcopyright{cc}
\setcctype{by}
\acmConference[ASE '26]{Proceedings of the 41st IEEE/ACM International Conference on Automated Software Engineering}{October 12--16, 2026}{Munich, Germany}
\acmBooktitle{Proceedings of the 41st IEEE/ACM International Conference on Automated Software Engineering (ASE '26), October 12--16, 2026, Munich, Germany}
\acmDOI{10.1145/3832783.3834386}
\acmISBN{979-8-4007-2882-2/2026/10}
\acmSubmissionID{ase26main-p1373-p}
\received{2026-03-26}
\received[accepted]{2026-06-18}

\definecolor{lavendergrey}{HTML}{8497B0}
\definecolor{powerblush}{HTML}{F7A399}
\definecolor{teagreen}{HTML}{B4C77F}
\definecolor{lightteagreen}{HTML}{DEE6C6}
\definecolor{vintageberry}{HTML}{883955}
\newcommand{\secr}[1]{\textcolor{lavendergrey}{#1}}
\newcommand{\excr}[1]{\textcolor{powerblush}{#1}}

\definecolor{poneonly}{HTML}{B4C77F}
\definecolor{mixed}{HTML}{121A40}
\definecolor{ptwoonly}{HTML}{F7B32B}

\newtcolorbox{mybox}{
  enhanced,
  colback=teagreen!5!white,
  colframe=lightteagreen!75!black,
  coltitle=black,
  fonttitle=\bfseries,
  title=Insight,
  boxsep=0.6mm,
  left=0.6mm,
  right=0.6mm,
  top=0.6mm,
  bottom=0.6mm,
  attach boxed title to top left={xshift=2mm,yshift*=-\tcboxedtitleheight/2},
  boxed title style={
    colback=white,
    colframe=white,
    boxrule=0pt,
    sharp corners
  }
}

\begin{document}

\begin{abstract}
AI coding assistants are increasingly embedded in software engineering practice, yet the literature still offers no cumulative, easily interpretable picture of their effects on software developer productivity. The central problem is not a lack of studies, but a lack of comparability: prior work operationalizes productivity through heterogeneous markers, settings, and reporting styles, making cross-study conclusions difficult. To address this, we conduct a staged systematic literature review. In the first stage, we synthesize studies that explicitly examine developer productivity. In the second, we extend the scope to productivity-related developer outcomes using SPACE as a theory-informed lens, capturing human-centered evidence that explicit productivity studies often omit. Across both stages, we analyze how productivity is operationalized, which confounding factors are discussed, and how reported effects are distributed.

We identify 187 unique productivity and productivity-related markers, while individual studies typically only use five, revealing a broad but highly fragmented evidence base. Explicit productivity research is dominated by time-, output-, and artifact-oriented proxies, whereas the broader SPACE-aligned corpus makes satisfaction, psychological aspects, and well-being substantially more visible. Reported effects are often positive, but this pattern is stronger in perceived and tendency-based findings than in objective or statistically evaluated ones. Overall, the literature suggests that AI coding assistants can support productivity, but current evidence does not yet support simple, context-independent conclusions. The key contribution of this review is therefore a structured basis for making productivity evidence more interpretable, multidimensional, and comparable across studies. 
\end{abstract}

\maketitle

\section{Introduction}

Coding assistants such as ChatGPT and GitHub Copilot are rapidly reshaping software engineering practice across tasks such as code and test generation\cite{taking_flight_with_copilot,navigating_the_complexity_of_generative_ai_adoption_in_software_engineering,codecommentgen,hao2024empiricalstudydevelopersshared,chatgpt_incorrectness_detection_in_software_reviews,aicoding_assistants_in_practice}, code review\cite{codereview,hao2024empiricalstudydevelopersshared,accountabilityincodereview,chatgpt_incorrectness_detection_in_software_reviews,codereviewlitrev}, and automated program repair\cite{automatedprogramrepair,contrastrepair}. By 2025, 84\,\% of professional developers reported that they either want to adopt or have already adopted AI coding assistants in their workflow\cite{StackOverflow2025}, making productivity effects a central practical concern.

Despite the growing number of studies on AI coding assistants, the evidence on developer productivity remains difficult to interpret. 
Reported effects vary substantially across studies: some find large time savings when completing a task\cite{the_impact_of_ai_on_developer_productivity}, no statistically significant time savings in some phases\cite{echoes_of_ai}, and some even slower task completion\cite{measuring_the_impact_of_early_2025_ai_on_experienced_open_source_developer_productivity}. 
These differences are not surprising: studies examine different kinds of tasks (researcher-designed exercises\cite{the_impact_of_ai_on_developer_productivity}, realistic maintenance tasks\cite{echoes_of_ai}, and real repository issues\cite{measuring_the_impact_of_early_2025_ai_on_experienced_open_source_developer_productivity}) and use highly heterogeneous productivity measures. 
Across the literature, “productivity” may refer to task completion time\cite{ai_coding_assistants}, developer perceptions\cite{harnessing_the_potential_of_gen_ai_coding_assistants_in_public_sector_software_development,will_i_be_replaced,ai_coding_assistants,the_space_of_ai}, repository activity metrics (e.g., commits, pull requests, lines of code)\cite{using_llm_coding_assistants_to_increase_software_development_productivity_by_33,ai_assisted_programming_may_decrease_the_productivity_of_experienced_developers_by_increasing_maintenance_burden,generative_ai_and_labour_productivity}, or observational findings\cite{ai_coding_assistants,beyond_code_generation}.
Even task completion time itself is operationalized in multiple ways (e.g., direct measurement, Likert-scale perceptions, percentage estimates, or qualitative interview results). As a result, findings are hard to compare, and the field still lacks a coherent view of when and how AI coding assistants improve productivity.

This comparability problem is not merely a methodological inconvenience, but directly affects the conclusions that researchers and practitioners draw from the literature. 
When results from fundamentally different study settings are interpreted side by side as if they were equivalent, the field risks overgeneralizing effects that may depend on task type, population, study environment, or unreported contextual factors.  
The consequences are twofold: practitioners may base adoption decisions on evidence from study designs that do not reflect realistic software engineering work, and researchers may repeatedly reproduce similar weaknesses because the literature lacks a structured overview of which findings are comparable, under what conditions, and with which confounders in mind. 
The key question for a review is therefore not only what effects have been reported, but whether existing syntheses make those effects interpretable and comparable in the first place.

Existing literature reviews provide useful orientation, but do not yet make productivity evidence interpretable and comparable across heterogeneous contexts\cite{ferino_novice_2025,mohamed_impact_2025,hou_large_2024,viet_large_2024,friso_usa_ai-powered_2024}. Prior reviews either take a broad LLM-in-software-engineering perspective\cite{hou_large_2024,viet_large_2024}, focus on specific populations\cite{ferino_novice_2025}, or summarize effects without systematically tracing operationalizations, study conditions, and contextual factors relevant to comparability\cite{mohamed_impact_2025}. To our knowledge, no existing secondary study jointly analyzes these dimensions to make productivity evidence comparable and interpretable.

We address this gap with a staged systematic literature review (SLR) designed to improve the interpretability and comparability of evidence on AI coding assistants' effects on software developers' productivity. 
The staged design reflects two analytically distinct synthesis objectives. 
Explicit productivity studies provide the clearest basis for analyzing how productivity is defined, measured, reported, and under which conditions effects emerge.
At the same time, productivity is widely understood as multidimensional, and prior work does not consistently label all developer-centered, productivity-relevant outcomes as productivity.
We therefore structure the review in two stages: the first synthesizes explicit productivity studies, while the second synthesizes studies reporting outcomes aligned with one or more dimensions of the SPACE framework (Satisfaction and well-being, Performance, Activity, Communication and collaboration, and Efficiency and flow) — even when productivity is not the primary stated construct.  
We do not assume prior studies consistently apply SPACE or map cleanly to its dimensions.
Instead, we use it as a theory-informed scoping lens, integrating broader human-centered evidence without collapsing into the explicit-productivity corpus. 
This staged design supports both a precise analysis of direct productivity claims and a broader account of productivity-related effects, reducing construct undercoverage and strengthening the basis for cumulative research and evidence-informed practice.

Our analysis produced a taxonomy that organizes the methodological fragmentation, arranging research insights according to their operationalization approaches. 
From this taxonomy, we derive three key observations: (i) human-centered dimensions (e.g., developers' well-being) are rarely investigated in explicit productivity research; (ii) confounders are not systematically analyzed; and (iii) opinion- and tendency-based findings are largely positive, whereas statistically evaluated and objective results contain substantially more neutral and negative findings.
This strengthens the observation that the central challenge is not the absence of evidence, but the absence of an interpretable and comparable evidence base.
We contribute an important structured basis for making heterogeneous productivity findings more comparable and interpretable across software engineering contexts. 
Our consolidated overview of markers, confounders, and effect directions provides the shared ground the field needs to move from isolated studies toward a cumulative and coherent evidence base.
For academia, this supports more cumulative and better-calibrated research designs; for industry, it enables more defensible interpretation of productivity claims when evaluating AI coding assistants for real development environments.

To summarize, our contributions are as follows:

\begin{itemize}[topsep=5pt]
    \item We conduct a staged systematic literature review of AI coding assistants' impact on software developers' productivity(-related) markers. 
    \item We compile and categorize 187 unique productivity and productivity-related markers, identifying which aspects are measured frequently and which are largely missing.
    \item We provide a first list of confounding variables analyzed for results in explicit productivity research.
    \item We synthesize reported effects of coding assistant usage on explicit productivity markers and summarize where evidence is consistent, mixed, or absent across study contexts.
    \item We derive implications for researchers designing future evaluations and practitioners assessing productivity impact. 
\end{itemize}

\section{Background and Related Work}

\miniparagraph{Productivity in Software Engineering}
Developer productivity has been studied for decades without converging on a universal definition or metric set. 
Early approaches relied on output-oriented indicators such as lines of code\cite{jones_measuring_1978}, while recent work recognizes it as multi-faceted and not to be reduced to a single measure\cite{space,noda_devex_2023}.

This shift is reflected in multidimensional frameworks such as DevEx and SPACE. DevEx frames developer productivity through developer-centric drivers such as feedback loops, cognitive load, and flow state, and recommends combining subjective perceptions with objective process indicators\cite{noda_devex_2023}. SPACE similarly argues for a multidimensional view of productivity and distinguishes five dimensions: Satisfaction and well-being, Performance, Activity, Communication and collaboration, and Efficiency and flow\cite{space}. 
These frameworks are relevant for two reasons: they motivate why single-metric interpretations of AI coding assistant productivity gains are often incomplete, and they provide a lens for identifying productivity-related outcomes in studies that do not explicitly use the term \emph{productivity} while still reporting developer-centered effects relevant to productivity in practice.

\miniparagraph{Secondary Studies on AI Coding Assistants and Developer Productivity}
Existing review studies provide valuable orientation for this rapidly evolving research area, but do not yet deliver the synthesis required to make productivity evidence interpretable and comparable across heterogeneous contexts\cite{ferino_novice_2025,mohamed_impact_2025,hou_large_2024,viet_large_2024,friso_usa_ai-powered_2024}. 
Available reviews differ substantially in scope, population focus, source policy, and synthesis objective, making them informative for mapping the field, but less suitable for resolving the comparability problem that motivates our work.

Ferino et al.\cite{ferino_novice_2025}\ focus on novice developers’ adoption of LLMs, covering motivations, perceptions, and experiences in only educational and early-career contexts. 
Mohamed et al.\cite{mohamed_impact_2025}\ are the closest prior work, conducting an SLR on LLM-assistant productivity across 37 studies, mapping findings to SPACE. 
Our review differs in analytical purpose: where Mohamed et al.\ characterize study methods and synthesize benefits, risks, and SPACE-dimension coverage, we explicitly target the \emph{comparability} problem by analyzing marker-level operationalizations, extracting confounding variables, and preserving study-setting differences in effect synthesis. 
We also use SPACE, but with a different purpose: as a staged scoping mechanism to recover human-centered evidence that an explicit-productivity-only review would miss.

Further, we considered broader LLM-in-software-engineering reviews\cite{hou_large_2024,viet_large_2024}, which map application areas and trends but are not designed to synthesize human-centered productivity evidence across heterogeneous settings. We additionally examined Friso and Kalava\cite{friso_usa_ai-powered_2024}, but their scope and reporting style differ substantially from protocol-driven SLRs, limiting direct comparability.

The source policy also varies across prior reviews. 
Most existing SLRs are restricted to peer-reviewed studies\cite{mohamed_impact_2025,ferino_novice_2025,hou_large_2024}, a defensible choice for methodological rigor. 
However, for productivity impact in practice, influential evidence also appears in industry reports and preprints reporting original study designs and results. 
Excluding these improves uniformity but risks omitting practically relevant evidence. We therefore include peer-reviewed studies, industry reports, and preprints, but only when they report original empirical designs and results, and we label them explicitly in the analysis.

Taken together, to the best of our knowledge, no existing secondary study is explicitly designed to make AI coding assistant productivity evidence \emph{comparable and interpretable} across studies by jointly analyzing marker-level operationalizations, study settings and documented confounding variables.

\section{Methodology}

We conducted a \emph{staged} SLR to synthesize evidence on the impact of AI coding assistants on software development. 
The core challenge was obtaining interpretable results across heterogeneous study designs, tasks, and measurement approaches, making a statistical meta-analysis of effect sizes currently infeasible. 
Instead, we aimed for a structured and transparent synthesis of (i) how productivity is operationalized via markers used to assess productivity and productivity-related outcomes in human-centered studies of AI coding assistants, (ii) which confounders are discussed in relation to reported effects, and (iii) how reported effects are distributed across explicit productivity and productivity-related outcomes.

\subsection{Research Questions}

To support an interpretable synthesis across heterogeneous evidence, we first characterize how prior work operationalizes productivity and productivity-related outcomes, then extract confounding variables, and finally summarize reported effects at the marker level. We state the following research questions:

\begin{questions}
    \item 
    How is the productivity impact of AI coding assistants on software developers \textit{operationalized} and \textit{measured} in existing software engineering research regarding
    \begin{questions}
        \item explicit productivity, and
        \item productivity-related effects based on the SPACE dimensions?
    \end{questions}
    
    \item 
    Which confounding variables are discussed in relation to the reported explicit productivity effects?

    \item 
    What effects of AI coding assistants are reported for explicit productivity markers?
\end{questions}

\noindent
\ques{1} focuses on \emph{operationalization} (i.e., which markers are used and how they are measured), \ques{2} captures contextual interpretation and limitations (i.e., confounding influences explicitly discussed in the studies), and \ques{3} summarizes reported effect directions to enable cautious cross-study comparison under heterogeneous reporting.
Essentially, we cannot go directly to \ques{3} without a prior analysis and contextualization of how and under which constraints the results were obtained. This is a key and unique aspect of our analysis compared to other secondary studies on this topic.

We use the term \emph{marker} rather than \emph{metric} because many studies are not operationalized via standardized quantitative metrics, but instead with qualitative/mixed-methods, such as interviews. Thus, we define markers as study-reported operationalizations of higher-level constructs (e.g., productivity, satisfaction, activity), including quantitative measures and qualitative operationalizations.

\subsection{Review Design and Search Strategy}

We design a staged systematic literature review following the SIGSOFT Empirical Standard for Systematic Reviews\cite{sigsoft}. 
Phase one synthesizes studies explicitly examining productivity with AI coding assistants. 
Phase two extends this with a scoping stage using SPACE to capture productivity-related developer outcomes not labeled as productivity. 
The full review protocol, including detailed processes and author responsibilities can be found in the supplementary material.

\subsubsection{Staged Review Design}
We conducted the review in two stages corresponding to \ques{1.1} and \ques{1.2} with two complementary synthesis objectives. 
Phase one focuses on studies explicitly examining developer productivity with AI coding assistants (\secr{ICp1}), providing the most direct and methodologically interpretable corpus for analyzing how productivity is explicitly operationalized and grounds the phase-specific analyses of explicit productivity markers, confounders, and reported effects. 
Phase two focuses on productivity-related outcomes aligned with one or more SPACE dimensions (\secr{ICp2}), acknowledging that relevant developer-centered evidence is not always labeled as productivity.

This design was empirically confirmed during phase one screening: multiple studies reported outcomes conceptually aligned with SPACE dimensions without naming productivity as the primary construct, and would have been excluded under an explicit-productivity-only criterion. 
Phase one was retained as the explicit-productivity corpus and complemented by phase two as a theory-informed extension, with phase-specific analyses preserved throughout to broaden conceptual coverage without collapsing distinct forms of evidence.
Merging the two phases would therefore risk conflating methodologically distinct forms of evidence.

\subsubsection{Selection Criteria}

We defined inclusion and exclusion criteria to include only studies that provide direct, human-centered evidence relevant to software developers' work with AI coding assistants.
The key criteria are summarized as follows:

\begin{itemize}[leftmargin=0.2cm]
    \item \textbf{Phase-specific scope criteria:}
    \begin{itemize}[leftmargin=0.1cm]
        \item \textbf{\secr{ICp1} (Phase one, \ques{\textbf{1.1}}):} The paper explicitly examines productivity effects of coding assistants on software developers.
        \item \textbf{\secr{ICp2} (Phase two, \ques{\textbf{1.2}}):} The paper examines productivity-related effects aligned with one or more SPACE dimensions.
    \end{itemize}
    
    \item \textbf{Common inclusion criteria (both phases):}
    \begin{itemize}[leftmargin=0.1cm]
        \item Addresses a software engineering activity/developers' work (\secr{IC2}).
        \item Directly reports productivity(-related) results (\secr{IC3}).
        \item Involves human participants/data from human developers (\secr{IC4}).
        \item Publicly available in English (\secr{IC5}).
        \item Peer-reviewed, preprint, or industry report with an original empirical study design and results (\secr{IC6}).
    \end{itemize}
    
    \item \textbf{Key exclusions:}
    \begin{itemize}[leftmargin=0.1cm]
        \item Published before 2023, predating the LLM-assistant period (\excr{EC1}).
        \item Exclusively teaching/student/non-developer populations (\excr{EC2}).
        \item Fully automated benchmark/ground-truth evaluations and LLM--LLM comparisons (\excr{EC3}).
        \item Humans only for ground-truth creation/output evaluation (\excr{EC4}).
        \item Other gray literature not meeting \secr{IC6} (\excr{EC5}).
    \end{itemize}
\end{itemize}

\noindent
Mixed-participant studies (e.g., developers and students) are included when conducted in non-teaching contexts and labeled accordingly. The temporal scope starts in 2023 (\excr{EC1}), reflecting mainstream LLM availability from 2022 onward. Full criterion formulations are provided in the supplementary material.

\subsubsection{Paper Collection and Screening Procedure}

To reduce dependence on a single search source and ranking mechanism, we collected candidate papers from three complementary sources. 
\begin{enumerate}[label=(\alph*), leftmargin=*]
    \item a venue-based search in major software engineering venues (to cover established software engineering venues), 
    \item an automated keyword-based search using the Semantic Scholar API (for scalable retrieval across indexed sources), and
    \item a manual search using Google and Google Scholar (for targeted recovery of additional studies and industry reports).
\end{enumerate}

\miniparagraph{Venue-based search}
We evaluated research track papers of 12 major software engineering venues (including ASE, ICSE, and ESE; full list in supplementary). 
The venue-based screening covered \textbf{3,746} papers in total, resulting in \textbf{1} relevant paper in phase one (\secr{ICp1}) and additional \textbf{4} (5 total) relevant papers in phase two (\secr{ICp2}).

\miniparagraph{Semantic Scholar API search}
As a second source, we queried the Semantic Scholar API using 120 iteratively refined keyphrase combinations and collected up to 20 results per query. 
Keyphrases were 
constructed from template strings combining placeholders for person type (PERSON: \textit{human, participant, developer, practitioner, expert}) and tool type (CHAT: 
\textit{chatgpt, LLM, coding assistant, chatbot, copilot}), combined with fixed  keyphrases. 
Placeholders and combinations were refined iteratively, with unsuitable ones dropped. 
The template and static phrases are as follows:
\begin{itemize}[leftmargin=0.2cm]
    \item \textbf{Templates:}
    \begin{itemize}[leftmargin=0.1cm]
        \item PERSON using CHAT
        \item How CHAT impact PERSONs
        \item Do CHAT increase productivity of PERSONs?
        \item Productivity of CHAT in practice
        \item Software PERSON on CHAT-usage
        \item Impact of CHAT on Software PERSON
    \end{itemize}
    \item \textbf{Static:}
    \begin{itemize}[leftmargin=0.1cm]
        \item Developers using LLMs
        \item LLMs in development
        \item how do coding assistants improve productivity in practice
        \item Do coding assistants increase productivity of software developers?
        \item Developer productivity coding assistants
        \item Effectiveness LLMs in Software Engineering
        \item Coding Assistants in Software Development
        \item Chatbots in Software Development
    \end{itemize}
\end{itemize}
The full list of template and placeholder combinations is provided in the supplementary material.

After deduplication, this step contributed 893 unique candidate papers, contributing \textbf{13} relevant papers in each phase (\textbf{26} total).

\miniparagraph{Manual search}
As a third source, we manually queried \textbf{10} productive keyphrases in Google and Google Scholar, adapting syntax where needed (e.g., appending \texttt{pdf} to reduce blog results, or reformulating \textit{llm impact developers} as \textit{How LLMs impact developers}). 
Screening the first {20} results per query added {79} further unique candidate papers.
One candidate was provided by Google's AI summary.
This summary is generated automatically with most Google searches and is thus part of the search interface rather than an active AI tool used in our collection pipeline.
This source contributed \textbf{13} relevant papers in phase one and \textbf{1} in phase two (\textbf{14} total).

\paragraph{Screening and full-text verification}
We screened candidates by title and abstract against the selection criteria. 
All screening and verification steps were performed by the first author. 
The third author independently screened the first 1,000 venue-paper titles as a calibration subset. 
Disagreements were resolved through discussion. 
For any undecided paper, the second author was consulted.
For abstract screening, the second author reviewed venue paper abstracts for calibration.
When abstracts were ambiguous on the \emph{phase-specific scope criterion} (\secr{ICp1}/\secr{ICp2}), we inspected the full text for keywords related to (i) the AI tool (e.g., ChatGPT, LLM), (ii) the study population (e.g., human, participant), and (iii) the productivity(-related) construct under consideration.
Papers passing abstract screening underwent full-text verification using the selection criteria, with particular attention to participant population and the requirement that productivity(-related) results are reported rather than only intermediate measures.
For any paper where the first author remained undecided, the second author was consulted.

We defined a control set of 10 papers identified through a small-scale manual pre-study and used it to validate search completeness. Search expansion was stopped after the third source once all control papers had been recovered.

\subsection{Data Processing and Synthesis Procedure}
The first two authors aligned on coding decisions after each step; phase-specific differences are stated below.

\subsubsection{Step 1: Extracting Markers}

In phase one, we extracted all markers used to assess productivity, excluding markers unrelated to productivity (e.g., use-case reporting without productivity implications) even when reported in otherwise relevant papers.

In phase two, we applied the same extraction logic but evaluated whether each paper's objective or \ques{}s addressed developer impact aligned to one or more SPACE-related constructs, rather than explicit productivity.
We reassessed the phase-one pool for previously excluded markers that aligned with SPACE, allowing re-inclusion where appropriate. 
Derived analyses combining existing markers without introducing a new operationalization (e.g., correlations between previously extracted constructs) were excluded.

\miniparagraph{Operationalizing SPACE alignment (phase two)}
Because SPACE defines dimensions conceptually without formal classification criteria, we operationalized inclusion by examining each paper's stated objective, introduction, and (where applicable) research questions/ hypotheses to determine whether the investigated construct directly aligned with one or more SPACE dimensions from a developer-centered perspective. 
Mapping required clear conceptual correspondence to a SPACE dimension rather than indirect influence. 
Cases with conceptual overlap were resolved through author discussion.
For example, \citet{beyond_code_generation} investigate perceived usefulness and trust in a single \ques{}: usefulness was retained as its operationalization directly addresses SPACE dimensions (e.g., reducing repetitive tasks maps to \textit{Efficiency and Flow}), while trust, a perception \emph{of} the tool rather than a developer-centered outcome, was excluded.

\subsubsection{Step 2: Mapping Markers to Normalized Labels}

In both phases, we mapped each extracted marker to a normalized \emph{label} via open coding, enabling cross-study comparison across semantically similar but differently described operationalizations (e.g., varying formulations of time savings or code-activity measures). 
The output of this step is therefore a concept-level normalization layer above the raw study-reported markers.
Quantitative markers were generally straightforward to normalize, while qualitative ones required more semantic interpretation. To calibrate the coding scheme, the first two authors jointly categorized an initial subset of 75 markers in a card-sorting session.
Then, the first author completed the remainder and discussed ambiguous cases with the second author.

\subsubsection{Step 3: Grouping into Higher-Level Categories}

We grouped normalized concepts into higher-level categories inductively via open coding (phase one), extending the structure in phase two to accommodate SPACE-aligned markers not present previously. 
Categories were further grouped into generalizable groups for readability. 
An overview of all groups, categories, and markers can be found in our results in \autoref{fig:marker-tree}.

\subsubsection{Step 4: Collecting explicitly discussed confounding variables}

To operationalize \ques{2}, we collected confounding variables only when explicitly discussed in relation to extracted explicit productivity markers (phase one). 
For each marker, we identified variables explicitly linked to the reported effect and assigned them a normalized label, synthesizing \emph{discussed contextual influences} rather than all variables measured in the included studies.

\subsubsection{Step 5: Coding Reported Effects}

To operationalize \ques{3}, we coded the reported effect direction of each phase-one marker using an ordinal scheme: \emph{negative}, \emph{mixed}, \emph{positive}, \emph{neutral/no-effect}.
We introduced this harmonization layer because studies report similar concepts using heterogeneous measurement formats (e.g., time measurements, percentage changes, Likert-scale perceptions, or qualitative findings), making a quantitative comparison infeasible.

We proceeded three-ways, depending on reporting format. 
\textbf{(i) Statistical results:} we coded effect direction based on reported significance, with effect sizes extracted if available. 
\textbf{(ii) Proportion-based results:} we derived proportions of positive, neutral, and negative responses and coded the dominant category, using \emph{mixed} when two categories were within 5 percentage points. 
\textbf{(iii) Reported tendencies:} we coded direction based on tendencies stated by the authors. 
Effect directions were summarized per marker by counting occurrences; multiple results for the same marker within a paper were counted separately when derived from distinct data.

This aggregation inevitably abstracts from deeper analyses in the original studies and compares substantively different result types.
Nevertheless, it was necessary given the substantial heterogeneity in operationalizations, analytical depth, and reporting formats across studies. 
Without a shared comparison layer, the synthesis would risk conflating differences in reporting practice with differences in substantive findings. 
The directional coding is thus not a meta-analytic effect-size estimate, but a structured and methodologically defensible summary of reported tendencies. 
We also extracted evidence type per finding to distinguish quantitative metrics, perceptions, and qualitative findings. 

\section{Results}

We analyzed a total of 45 papers (27 in phase one and 18 in phase two), summarized in \autoref{tab:paper-overview}.
Overall, 26 of the 45 papers were peer reviewed (57.8\%), with a lower share in phase one than in phase two (37.0\% vs.\ 88.9\%).
The most common methods were surveys and experiments, or combinations thereof (\textit{Top Methods}).
Participant numbers ranged from 5 to 88022 (\textit{N (min--max)}), depending on the method used.
The smallest sample came from a think-aloud study embedded in a larger study setup\cite{taking_flight_with_copilot}, and the largest from an analysis of participant activity data in repositories\cite{beyond_code_the_multidimensional_impacts_of_large_language_models_in_software_development}.
Given this skewed distribution, we report participant ranges rather than averages.
Participants were predominantly software engineering professionals.
Study duration differed between phases.
Phase two papers were almost exclusively one-shot studies (17/18), whereas phase one included a broader spread of durations, including all long-term studies in the sample (6/27).
Interestingly, phase one papers used established productivity frameworks only rarely, with 5 papers using SPACE\cite{echoes_of_ai,harnessing_the_potential_of_gen_ai_coding_assistants_in_public_sector_software_development,the_space_of_ai,significant_productivity_gains_through_programming_with_large_language_models,taking_flight_with_copilot} and 3 papers using other frameworks\cite{beyond_code_generation,the_impact_of_generative_ai_coding_assistants_on_developers_who_are_visually_impaired,investigating_the_impact_of_ai_assisted_tools_on_software_practitioner_well_being}.

\begin{table}[htbp]
    \centering
    \small
    \caption{Overview of papers by phase}
    \label{tab:paper-overview}
    \begin{tabular}{lccc}
    \toprule
     & Phase One & Phase Two & Total \\
    \midrule\midrule
    Unique Paper & 27 & 18 & 45 \\
    Peer-Reviewed & 10 (37.04\%) & 16 (88.89\%) & 26 (57.78\%) \\
    Top Methods & \makecell[tc]{survey (16) \\ experiment (8)} & \makecell[tc]{survey (10) \\ experiment (4)} & \makecell[tc]{survey (26) \\ experiment (12)} \\
    N (min--max) & 5--88022 & 10--268 & 5--88022 \\
    Mixed & 2 (7.41\%) & 3 (16.67\%) & 5 (11.11\%) \\
    Study Length &  &  &  \\
    \hspace{5pt} one-shot & 17 (62.96\%) & 17 (94.44\%) & 34 (75.56\%) \\
    \hspace{5pt} short-term & 4 (14.81\%) & 1 (5.56\%) & 5 (11.11\%) \\
    \hspace{5pt} long-term & 6 (22.22\%) & 0 (0.00\%) & 6 (13.33\%) \\
    \bottomrule
    \end{tabular}
\end{table}

\subsection{\ques{1}: Markers of productivity research}

We report marker results phase-specific.
Phase one captures how prior work \emph{explicitly} operationalizes productivity and therefore provides the tighter basis for comparability.
Phase two is not intended as a second explicit-productivity corpus.
Instead, it focuses on developer-centered outcomes aligned with SPACE dimensions.
This approach highlights the aspects that are excluded by an explicit productivity-only lens.
We therefore interpret phase differences not simply as differences in marker counts, but as evidence of what explicit productivity research emphasizes versus what it overlooks.

\begin{table}[htbp]
    \centering
    \small
    \caption{Overview of markers per paper by phase}
    \label{tab:marker-overview}
    \begin{tabular}{p{4cm}ccc}
        \toprule
         & Phase One & Phase Two & Total \\
        \midrule\midrule
        Papers with markers & 27 & 31 & 45 \\
        Total Markers & 178 & 240 & 418 \\
        Unique Markers & 91 & 124 & 187 \\
        Unique Categories & 19 & 19 & 20 \\
        Median \# unique Markers per Paper & 5 & 5 & 6 \\
        Median \# Categories per Paper & 3 & 3 & 4 \\
        \bottomrule
    \end{tabular}
\end{table}

\autoref{tab:marker-overview} summarizes the extracted markers used to assess pro\-ductivity(-related) impacts of AI coding assistants on developers. Overall, we identified 187 unique markers. Both phases showed the same median of 5 unique markers per paper and a similar category distribution: each contained 19 unique categories, with a median of 3 categories per paper. This suggests a broad corpus-level evidence base but narrow study-level coverage: many markers exist, but each paper captures only a small part of the construct. To test whether the mix of peer-reviewed, industry, and preprint papers affected the identified markers, we used a PERMANOVA. 
We did not observe statistically detectable differences in marker distributions by peer-review status in either phase one ($p = 0.839$, pseudo-$F=0.874$, $R^2= 0.033$), or phase two ($p = 0.107$, pseudo-$F=1.168$, $R^2= 0.038$) paper pool, providing no evidence that marker distributions differ systematically by publication type.

\begin{mybox}{}
The field has collected a broad productivity evidence base, though the study-level coverage remains narrow.
Across the corpus 187 distinct markers are used, but individual studies usually only rely on a small subset, further limiting comparability.
\end{mybox}

\begin{figure*}[t]
    \centering
    \begin{forest}
    for tree={
      grow=east,
      parent anchor=east,
      child anchor=west,
      l sep=12mm,
      s sep=-1.6298mm,
      edge={line width=1.2pt},
      font=\small,
    },
    where level=0{text width=0mm}{},
    where level=1{text width=15mm}{},
    where level=2{text width=30mm}{},
    where level=3{text width=85mm}{},
    [
    [{Organizational and process marker\\(\textcolor{mixed}{128}, \textcolor{poneonly}{59}, \textcolor{ptwoonly}{69})}, edge={mixed, line width=1.4pt}
      [{Code authorship\\(\textcolor{mixed}{4}, \textcolor{poneonly}{3}, \textcolor{ptwoonly}{1})}, edge={mixed}
      ]
      [{Collaboration activity\\\raisebox{-0.25\height}{\includegraphics[height=0.35cm]{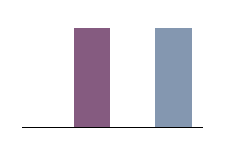}}(\textcolor{mixed}{10}, \textcolor{poneonly}{2}, \textcolor{ptwoonly}{8})}, edge={mixed}
      ]
      [{Documentation\\\raisebox{-0.25\height}{\includegraphics[height=0.35cm]{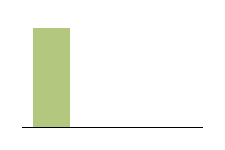}}(\textcolor{mixed}{11}, \textcolor{poneonly}{4}, \textcolor{ptwoonly}{7})}, edge={mixed}
      ]
      [{Organizational\\\raisebox{-0.25\height}{\includegraphics[height=0.35cm]{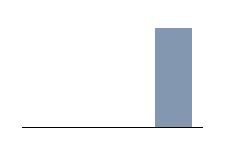}}(\textcolor{mixed}{2}, \textcolor{poneonly}{1}, \textcolor{ptwoonly}{1})}, edge={mixed}
      ]
      [{Task success and task quality\\\raisebox{-0.25\height}{\includegraphics[height=0.35cm]{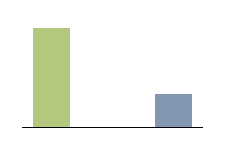}}(\textcolor{mixed}{17}, \textcolor{poneonly}{8}, \textcolor{ptwoonly}{9})}, edge={mixed}
      ]
      [{Time\\\raisebox{-0.25\height}{\includegraphics[height=0.35cm]{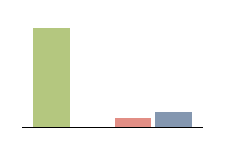}}(\textcolor{mixed}{84}, \textcolor{poneonly}{41}, \textcolor{ptwoonly}{43})}, edge={mixed}
        [\perceived{}  Coding speed\raisebox{-0.25\height}{\includegraphics[height=0.35cm]{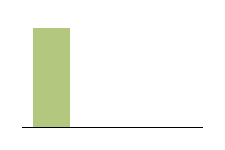}}\cite{will_i_be_replaced,ai_assisted_code_authoring_at_scale,study_of_software_developers_experience_using_the_github_copilot_tool_in_the_software_development_process}, edge={mixed}]
        [\perceived{}  Development speed\raisebox{-0.25\height}{\includegraphics[height=0.35cm]{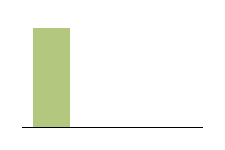}}\cite{the_impact_of_ai_tools_on_software_development,my_code_is_less_secure_with_gen_ai,youre_on_a_bicycle_with_a_little_motor,automated_code_review_in_practice,practices_and_challenges_of_using_github_copilot}, edge={mixed}]
        [\perceived{}  Efficiency\raisebox{-0.25\height}{\includegraphics[height=0.35cm]{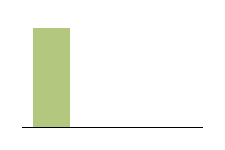}}\cite{navigating_the_complexity_of_generative_ai_adoption_in_software_engineering,llms_reshaping_of_people_processes_products_and_society_in_software_development,the_influence_of_using_llms_on_the_activities_of_a_software_testing_team,the_space_of_ai,significant_productivity_gains_through_programming_with_large_language_models,my_code_is_less_secure_with_gen_ai,developers_perceptions_on_the_impact_of_chatgpt_in_software_development}, edge={mixed}]
        [\perceivedobjective{}  Task completion time\raisebox{-0.25\height}{\includegraphics[height=0.35cm]{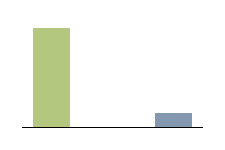}}\cite{measuring_the_impact_of_early_2025_ai_on_experienced_open_source_developer_productivity,the_impact_of_ai_on_developer_productivity,the_impact_of_ai_pair_programmers_on_code_quality_and_developer_satisfaction,echoes_of_ai,navigating_the_complexity_of_generative_ai_adoption_in_software_engineering,using_an_llm_to_help_with_code_understanding,ai_coding_assistants,harnessing_the_potential_of_gen_ai_coding_assistants_in_public_sector_software_development,significant_productivity_gains_through_programming_with_large_language_models,an_empirical_study_of_adoption_of_chatgpt_for_bug_fixing_among_professional_developers}, edge={mixed}]
        [\perceived{}  Time saving\raisebox{-0.25\height}{\includegraphics[height=0.35cm]{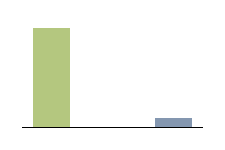}}\cite{llms_reshaping_of_people_processes_products_and_society_in_software_development,harnessing_the_potential_of_gen_ai_coding_assistants_in_public_sector_software_development,echoes_of_ai,the_impact_of_ai_tools_on_software_development,examining_the_use_and_impact_of_an_ai_code_assistant_on_developer_productivity_and_experience_in_the_enterprise,chatgpt_incorrectness_detection_in_software_reviews,ai_assisted_code_authoring_at_scale,the_impact_of_ai_pair_programmers_on_code_quality_and_developer_satisfaction,beyond_code_generation,navigating_the_complexity_of_generative_ai_adoption_in_software_engineering,study_of_software_developers_experience_using_the_github_copilot_tool_in_the_software_development_process,using_an_llm_to_help_with_code_understanding,impact_of_llms_on_team_collaboration_in_software_development}, edge={mixed}]
        [\perceivedobjective{}  Time spent on specific task/action\raisebox{-0.25\height}{\includegraphics[height=0.35cm]{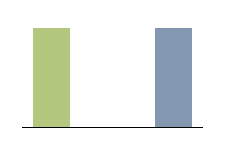}}\cite{my_code_is_less_secure_with_gen_ai,harnessing_the_potential_of_gen_ai_coding_assistants_in_public_sector_software_development,taking_flight_with_copilot,developers_perceptions_on_the_impact_of_chatgpt_in_software_development,impact_of_llms_on_team_collaboration_in_software_development}, edge={mixed}]
        [\perceived{}  Ability to meet deadline\cite{the_impact_of_ai_tools_on_software_development}, edge={ptwoonly}]
        [\perceived{}  AI-driven time pressure\cite{investigating_the_impact_of_ai_assisted_tools_on_software_practitioner_well_being}, edge={ptwoonly}]
        [\perceived{}  Coding efficiency\cite{the_impact_of_generative_ai_coding_assistants_on_developers_who_are_visually_impaired}, edge={ptwoonly}]
        [\perceived{}  Overtime due to AI tool usage\cite{investigating_the_impact_of_ai_assisted_tools_on_software_practitioner_well_being}, edge={ptwoonly}]
        [\perceived{}  Temporal demand\cite{using_an_llm_to_help_with_code_understanding}, edge={ptwoonly}]
        [\perceived{}  Time spent on on specific task/action\cite{taking_flight_with_copilot}, edge={ptwoonly}]
        [\objective{}  Time until PR closed\cite{automated_code_review_in_practice}, edge={ptwoonly}]
        [\objective{}  \# of characters per minute by developer\raisebox{-0.25\height}{\includegraphics[height=0.35cm]{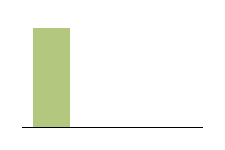}}\cite{significant_productivity_gains_through_programming_with_large_language_models}, edge={poneonly}]
        [\objective{}  \# tasks per day\raisebox{-0.25\height}{\includegraphics[height=0.35cm]{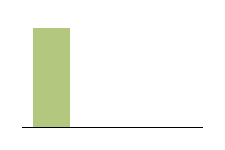}}\cite{the_ai_productivity_paradox}, edge={poneonly}]
        [\objective{}  Mean time to resolve\raisebox{-0.25\height}{\includegraphics[height=0.35cm]{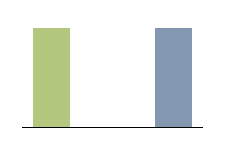}}\cite{security_copilot,the_ai_productivity_paradox}, edge={poneonly}]
        [\objective{}  PR cycle time\raisebox{-0.25\height}{\includegraphics[height=0.35cm]{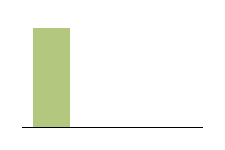}}\cite{intuition_to_evidence}, edge={poneonly}]
        [\objective{}  Requirement completion per minute\raisebox{-0.25\height}{\includegraphics[height=0.35cm]{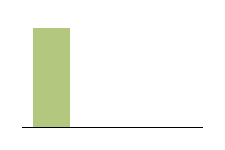}}\cite{significant_productivity_gains_through_programming_with_large_language_models}, edge={poneonly}]
        [\objective{}  Time until PR is reviewed / approved\raisebox{-0.25\height}{\includegraphics[height=0.35cm]{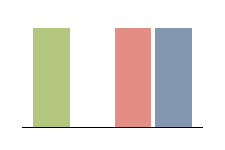}}\cite{the_effects_of_generative_ai_on_high_skilled_work,the_ai_productivity_paradox,intuition_to_evidence}, edge={poneonly}]
        [\objective{}  Time until PR merged\raisebox{-0.25\height}{\includegraphics[height=0.35cm]{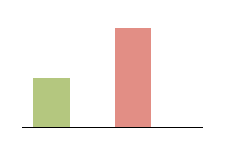}}\cite{the_impact_of_generative_ai_on_collaborative_open_source_software_development}, edge={poneonly}]
      ]
    ]
    [{Artifact and repository marker\\(\textcolor{mixed}{94}, \textcolor{poneonly}{52}, \textcolor{ptwoonly}{42})}, edge={mixed, line width=1.4pt}
      [{CI activity\\\raisebox{-0.25\height}{\includegraphics[height=0.35cm]{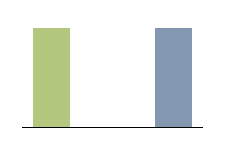}}(\textcolor{mixed}{3}, \textcolor{poneonly}{2}, \textcolor{ptwoonly}{1})}, edge={mixed}
      ]
      [{Code quality\\\raisebox{-0.25\height}{\includegraphics[height=0.35cm]{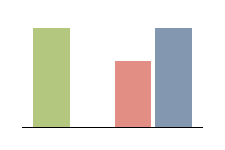}}(\textcolor{mixed}{43}, \textcolor{poneonly}{16}, \textcolor{ptwoonly}{27})}, edge={mixed}
      ]
      [{Code volume\\\raisebox{-0.25\height}{\includegraphics[height=0.35cm]{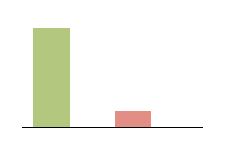}}(\textcolor{mixed}{8}, \textcolor{poneonly}{7}, \textcolor{ptwoonly}{1})}, edge={mixed}
      ]
      [{Repository activity\\\raisebox{-0.25\height}{\includegraphics[height=0.35cm]{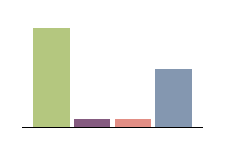}}(\textcolor{mixed}{25}, \textcolor{poneonly}{21}, \textcolor{ptwoonly}{4})}, edge={mixed}
      ]
      [{Security\\\raisebox{-0.25\height}{\includegraphics[height=0.35cm]{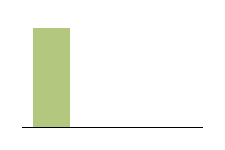}}(\textcolor{mixed}{12}, \textcolor{poneonly}{3}, \textcolor{ptwoonly}{9})}, edge={mixed}
      ]
      [{Production activity\\\raisebox{-0.25\height}{\includegraphics[height=0.35cm]{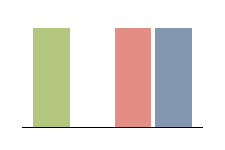}}(\textcolor{mixed}{3}, \textcolor{poneonly}{3}, \textcolor{ptwoonly}{0})}, edge={poneonly}
      ]
    ]
    [{AI marker\\(\textcolor{mixed}{87}, \textcolor{poneonly}{23}, \textcolor{ptwoonly}{64})}, edge={mixed, line width=1.4pt}
      [{AI assessment\\\raisebox{-0.25\height}{\includegraphics[height=0.35cm]{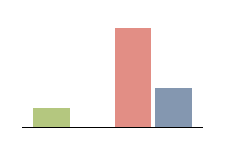}}(\textcolor{mixed}{60}, \textcolor{poneonly}{8}, \textcolor{ptwoonly}{52})}, edge={mixed}
        [\perceived{}  AI tool accuracy\raisebox{-0.25\height}{\includegraphics[height=0.35cm]{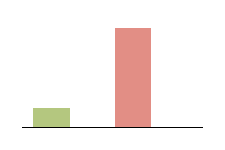}}\cite{llms_reshaping_of_people_processes_products_and_society_in_software_development,harnessing_the_potential_of_gen_ai_coding_assistants_in_public_sector_software_development,the_influence_of_using_llms_on_the_activities_of_a_software_testing_team,significant_productivity_gains_through_programming_with_large_language_models,taking_flight_with_copilot,will_i_be_replaced,ai_assisted_code_authoring_at_scale,the_impact_of_ai_tools_on_software_development,developers_perspective_on_todays_and_tomorrows_programming_tool_assistance,youre_on_a_bicycle_with_a_little_motor,transforming_software_development_with_generative_ai,practices_and_challenges_of_using_github_copilot,developers_perceptions_on_the_impact_of_chatgpt_in_software_development,using_an_llm_to_help_with_code_understanding}, edge={mixed}]
        [\perceived{}  Ability to identify issues found by AI alone\cite{automated_code_review_in_practice}, edge={ptwoonly}]
        [\perceived{}  AI code understandability\cite{practices_and_challenges_of_using_github_copilot}, edge={ptwoonly}]
        [\perceived{}  AI tool code quality\cite{practices_and_challenges_of_using_github_copilot}, edge={ptwoonly}]
        [\perceived{}  AI tool limitation\cite{chatgpt_incorrectness_detection_in_software_reviews,will_i_be_replaced,the_impact_of_ai_tools_on_software_development,the_impact_of_ai_pair_programmers_on_code_quality_and_developer_satisfaction,beyond_code_generation,navigating_the_complexity_of_generative_ai_adoption_in_software_engineering,practices_and_challenges_of_using_github_copilot,harnessing_the_potential_of_gen_ai_coding_assistants_in_public_sector_software_development}, edge={ptwoonly}]
        [\perceived{}  Assistance in bugfixing\cite{automated_code_review_in_practice}, edge={ptwoonly}]
        [\perceived{}  Assistance in code understandability\cite{navigating_the_complexity_of_generative_ai_adoption_in_software_engineering}, edge={ptwoonly}]
        [\perceived{}  Importance of AI-found issues\cite{automated_code_review_in_practice}, edge={ptwoonly}]
        [\perceived{}  Importance to have access to coding assistants\cite{ai_coding_assistants}, edge={ptwoonly}]
        [\perceived{}  LLM quality\cite{harnessing_the_power_of_llms,automated_code_review_in_practice,developers_perceptions_on_the_impact_of_chatgpt_in_software_development}, edge={ptwoonly}]
        [\perceived{}  Opinion of AI tool\cite{chatgpt_incorrectness_detection_in_software_reviews,ai_assisted_code_authoring_at_scale}, edge={ptwoonly}]
        [\perceived{}  Sentiment on AI tools\cite{why_do_developers_engage_with_chatgpt_in_issue_tracker}, edge={ptwoonly}]
        [\perceived{}  Team AI tool learning curve\cite{impact_of_llms_on_team_collaboration_in_software_development}, edge={ptwoonly}]
        [\perceived{}  Usefulness\cite{developers_perspective_on_todays_and_tomorrows_programming_tool_assistance,practices_and_challenges_of_using_github_copilot,using_an_llm_to_help_with_code_understanding}, edge={ptwoonly}]
        [\perceived{}  AI tool quality\raisebox{-0.25\height}{\includegraphics[height=0.35cm]{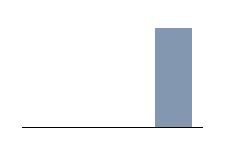}}\cite{examining_the_use_and_impact_of_an_ai_code_assistant_on_developer_productivity_and_experience_in_the_enterprise,significant_productivity_gains_through_programming_with_large_language_models}, edge={poneonly}]
      ]
      [{AI tooling\\\raisebox{-0.25\height}{\includegraphics[height=0.35cm]{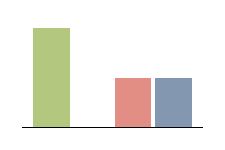}}(\textcolor{mixed}{8}, \textcolor{poneonly}{4}, \textcolor{ptwoonly}{4})}, edge={mixed}
      ]
      [{AI usage\\\raisebox{-0.25\height}{\includegraphics[height=0.35cm]{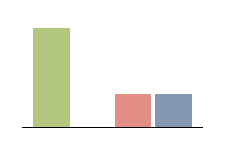}}(\textcolor{mixed}{19}, \textcolor{poneonly}{11}, \textcolor{ptwoonly}{8})}, edge={mixed}
      ]
    ]
    [{Human marker\\(\textcolor{mixed}{80}, \textcolor{poneonly}{17}, \textcolor{ptwoonly}{63})}, edge={mixed, line width=1.4pt}
      [{Personal career aspects\\\raisebox{-0.25\height}{\includegraphics[height=0.35cm]{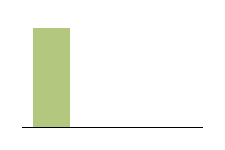}}(\textcolor{mixed}{14}, \textcolor{poneonly}{1}, \textcolor{ptwoonly}{13})}, edge={mixed}
      ]
      [{Psychological aspects\\\raisebox{-0.25\height}{\includegraphics[height=0.35cm]{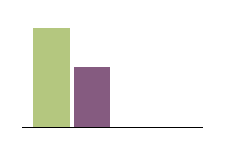}}(\textcolor{mixed}{27}, \textcolor{poneonly}{8}, \textcolor{ptwoonly}{19})}, edge={mixed}
      ]
      [{Satisfaction\\\raisebox{-0.25\height}{\includegraphics[height=0.35cm]{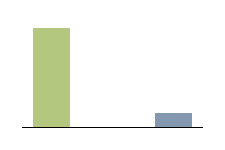}}(\textcolor{mixed}{30}, \textcolor{poneonly}{8}, \textcolor{ptwoonly}{22})}, edge={mixed}
      ]
      [{Well-being\\(\textcolor{mixed}{9}, \textcolor{poneonly}{0}, \textcolor{ptwoonly}{9})}, edge={ptwoonly}
        [\perceived{}  Frustration\cite{using_an_llm_to_help_with_code_understanding}, edge={ptwoonly}]
        [\perceived{}  Guilt\cite{taking_flight_with_copilot}, edge={ptwoonly}]
        [\perceived{}  Job-related stress\cite{investigating_the_impact_of_ai_assisted_tools_on_software_practitioner_well_being}, edge={ptwoonly}]
        [\perceived{}  Mental fatigue\cite{investigating_the_impact_of_ai_assisted_tools_on_software_practitioner_well_being}, edge={ptwoonly}]
        [\perceived{}  Mental health impact\cite{investigating_the_impact_of_ai_assisted_tools_on_software_practitioner_well_being}, edge={ptwoonly}]
        [\perceived{}  Physical demand\cite{using_an_llm_to_help_with_code_understanding}, edge={ptwoonly}]
        [\perceived{}  Physical discomfort\cite{investigating_the_impact_of_ai_assisted_tools_on_software_practitioner_well_being}, edge={ptwoonly}]
        [\perceived{}  Stress\cite{investigating_the_impact_of_ai_assisted_tools_on_software_practitioner_well_being}, edge={ptwoonly}]
        [\perceived{}  Well-being\cite{investigating_the_impact_of_ai_assisted_tools_on_software_practitioner_well_being}, edge={ptwoonly}]
      ]
    ]
    [{No group\\(\textcolor{mixed}{29}, \textcolor{poneonly}{27}, \textcolor{ptwoonly}{2})}, edge={mixed, line width=1.4pt}
      [{Direct productivity assessment\\\raisebox{-0.25\height}{\includegraphics[height=0.35cm]{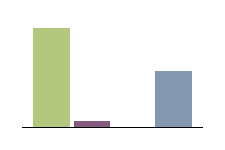}}(\textcolor{mixed}{29}, \textcolor{poneonly}{27}, \textcolor{ptwoonly}{2})}, edge={mixed}
      ]
    ]
    ]
    \end{forest}
    \caption{Markers used in productivity research. \textcolor{poneonly}{Green} lines indicate phase-one-only markers, categories, and groups; \textcolor{ptwoonly}{yellow} lines indicate phase-two-only; \textcolor{mixed}{black} lines indicate those found in both phases. Numbers below categories and groups show total collected (not unique) markers by phase. Bar graphs show the distribution of positive (green), mixed (purple), negative (red), and neutral/no-effect (gray) results for phase one markers. Leaf labels indicate whether operationalizations are perception-based (\perceived{}), objective (\objective{}), or both (\perceivedobjective{}). Perceived and objective markers are counted separately in the statistics.}
    \label{fig:marker-tree}
    \Description{This figure categorizes markers found in the literature into a hierarchy, showing how frequently each was studied, in which direction effects were reported, and into which category they belong, visualized as a tree. The tree structure displays 4 group labels (human marker, AI marker, artifact and repository marker, organizational and process marker) on level one, branching into 20 categories which then branch into the leaves which represent the found markers. 
    Each leaf shows the marker, whether it is an objective or perception-based marker, and, if the marker is from phase one, a bar graph showing the distribution of the effect directions per marker, and a citation to the paper(s) where the markers are found. Only three categories have their markers visible: Time, AI assessment, Well-being. For each category, there is also an aggregated bar chart representing all directions of the subsequent markers, and numbers representing the total occurrences, total in phase one, and total in phase two.}
  \end{figure*}

\paragraph{Markers and groups}
We identified 187 unique markers, organized into 20 categories and four groups in a tree-level structure.
Quantitative evidence was usually easier to consolidate than qualitative evidence: closely related metrics such as ``volume of code changes'' and ``lines of code added'' were unified under ``\#LoC'', whereas qualitatively phrased findings about faster work or time savings had to be consolidated at the level of shared meaning.
When the same construct appeared across multiple task contexts, we retained a single shared marker to avoid unnecessary fragmentation.
The resulting tree is visualized in \autoref{fig:marker-tree}.
Only a subset of categories is shown in detail, as displaying every marker would exceed the paper's space.
All markers with categories and their corresponding references are provided in the supplementary material.

We identified four main groups of markers: \emph{AI markers}, covering the AI tool itself; \emph{artifact and repository markers}, covering produced code and repository behavior; \emph{organizational and process markers}, covering task, workflow, and organizational outcomes; and \emph{human markers}, covering developers' experiences, perceptions, and well-being.
The most frequently addressed categories are \emph{Time} (84 total markers used), followed by \emph{AI assessment} (60) and \emph{Code quality} (43).
Because \emph{Time} belongs to the organizational and process marker group, this group is also the largest overall.

Two patterns are particularly notable.
First, both phases remain similarly narrow at paper level, as reflected in the identical medians.
Second, they differ systematically in \emph{what} they focus on.
Phase one contains more artifact and repository markers, whereas phase two is centered more around human and AI markers, and to a lesser extent organizational and process markers.
This contrast is also visible in the phase-unique categories: \emph{production activity} (artifact and repository), appears only in phase one, whereas \emph{well-being} (human) appears only in phase two.
Phase two therefore does not simply add markers, but captures forms of developer-centered evidence that explicit productivity studies largely leave outside the construct.

At paper level, the most frequent category pairings in phase one are \emph{Direct productivity assessment} with \emph{Time} (9 papers), followed by \emph{Task success and task quality} with \emph{Time}, and \emph{Time} with itself (7 papers each).
Self-combinations are common overall, such as \emph{Code quality} and \emph{Repository activity} (5 and 4 papers, respectively).
In phase two, the most frequent pairing is \emph{AI assessment} with \emph{Time} (13 papers), followed by self-combinations of \emph{Time} and \emph{AI assessment} (12 and 11 papers, respectively).
These pairing patterns highlight that studies often focus on narrow clusters of closely related markers.
Such obtained insights are therefore often narrow in scope even when the broader corpus appears rich.

\begin{mybox}{}
The phase contrast is a central result. 
Phase one shows the field's default productivity lens, centered on observable work-performance proxies and phase two shows the human and experiential dimensions that this default lens misses.
\end{mybox}

\miniparagraph{Human markers}
This group covers the developers' individual experiences, including \emph{Well-being}, \emph{Satisfaction}, \emph{Psychological aspects}, and \emph{Personal career aspects}, with \emph{Satisfaction} being the most frequent category.
Of the four main groups, it has the fewest occurrences and is concentrated in the phase two paper pool.
Markers related to \emph{Psychological aspects} and \emph{Satisfaction} roughly double or triple in phase two compared to phase one, while \emph{Personal career aspects} and \emph{Well-being} occur almost exclusively in phase two.
For instance, \emph{Well-being} is defined by three papers and captured exclusively through perceived markers including stress, physical and mental discomfort, and fatigue.
Human markers are therefore where the value of phase two becomes most visible: explicit productivity research includes these outcomes only sparsely, whereas the broader SPACE-aligned set shows that substantial productivity-relevant evidence exists outside explicit productivity labels.

\miniparagraph{AI markers}
This group describes telemetry and metadata on the interactions of developers and tools measured primarily objectively (\emph{AI usage}; e.g., usage frequency or amount of accepted code), as well as developers' perceptions of tool interaction at a more practical level (\emph{AI tooling}; e.g., perceived ease of use and integration challenges).
These categories are relatively balanced between the two phases.
The most prominent category is \emph{AI assessment}, which captures how developers evaluate the tool itself and can therefore shape both performance and perceptions of productivity (e.g., perceived suggestion accuracy).
Overall, \emph{AI assessment} accounts for 68\,\% (60/87) of all AI markers.
While AI markers are more frequent than human markers, they remain the second smallest group overall.
Their stronger phase-two presence is driven primarily by the concentration of \emph{AI assessment} markers in that phase.

\miniparagraph{Artifact and repository markers}
This group is the second largest in our paper pool and the only group for which phase one contains more markers than phase two.
It contains six distinct categories, although \emph{Production activity} (phase-one exclusive) and \emph{CI activity} contain only three markers each, drawn from one and two papers, respectively.
The largest category is \emph{Code quality}, collected in both phases but more often in phase two, and spanning 29 unique markers.
It includes objective markers such as cyclomatic complexity and test coverage as well as perceived markers such as perceived readability and understandability.
A notable subset consists of bug- and issue-related operationalizations and raw counts.
Because this group primarily focuses on code and repository artifacts, its markers are predominantly objective, with three categories being entirely objective (\emph{Production activity}, \emph{Repository activity}, and \emph{Code volume}).
Specifically, these markers are often operationalized through directly observable outcomes (e.g., \#PRs, \#commits).
The remaining categories are more mixed, but even there many perceived markers are phase-two exclusives.
Taken together, artifact and repository markers represent the clearest expression of the explicit-productivity lens foregrounded in phase one.

\miniparagraph{Organizational and process markers}
This is the largest group overall and contains six categories, all present in both phases.
The category sizes differ widely, with \emph{Code authorship} and \emph{Organizational} containing only two unique markers each (e.g., amount of AI code in repository and \emph{Company impact}, respectively).
\emph{Collaboration activity} captures the social aspects of organizational processes (e.g., satisfaction with communication flows) and occurs primarily in phase two, whereas the other categories are distributed more evenly.
Similar to artifact and repository markers, this group is objective-heavy, but some of its most central categories include both objective and perceived operationalizations.

The \emph{Time} category alone accounts for 65\,\% (84/128) of all organizational and process markers and is distributed evenly across phases, with as many markers based on objective measurements as on participants' perceptions.
Two patterns are particularly notable, both of which generalize beyond \emph{Time}.
First, some markers are closely related but still not directly comparable.
For example, the time until a pull request (PR) is merged is closely related to the time until a PR is reviewed or approved, but not identical, as merging follows approval and may occur later.
Second, the operationalization of even central constructs varies strongly across studies, making direct comparison difficult.
Task completion time, for example, has been measured directly\cite{measuring_the_impact_of_early_2025_ai_on_experienced_open_source_developer_productivity,the_impact_of_ai_pair_programmers_on_code_quality_and_developer_satisfaction,the_impact_of_ai_on_developer_productivity,significant_productivity_gains_through_programming_with_large_language_models,echoes_of_ai,using_an_llm_to_help_with_code_understanding} and reported in seconds, minutes, hours, or percentage changes; it has also been captured through perceived measures, including Likert-scale items\cite{echoes_of_ai,the_space_of_ai,aicoding_assistants_in_practice}, other closed-response formats\cite{navigating_the_complexity_of_generative_ai_adoption_in_software_engineering,harnessing_the_potential_of_gen_ai_coding_assistants_in_public_sector_software_development}, and qualitative coding\cite{an_empirical_study_of_adoption_of_chatgpt_for_bug_fixing_among_professional_developers}.
These measures were applied at the level of tasks or subtasks, in both artificial and realistic settings, and with different forms of evidence and reporting.
As a result, even clearly related findings can usually only be compared at the level of direction, such as whether task completion time improved or worsened, but not by degree.

\begin{mybox}{}
The large number of markers should not be read as measurement richness alone. It also signals a fragmented operationalization landscape in which even highly central constructs such as \emph{Time} are often only comparable at the level of direction, not magnitude.
\end{mybox}

\miniparagraph{Direct productivity assessment}
We define \emph{Direct productivity assessment} as a category of its own because all other markers represent proxy operationalizations, whereas this category captures productivity explicitly from developers through instruments such as Likert-scale items, multi-item questionnaires, and qualitative coding.
With 29 total occurrences, it is one of the more common categories.
At the same time, these measures vary substantially in wording and scale design.
Many Likert-based formulations, not limited to this category, do not permit a clear negative interpretation: disagreement with ``My productivity is improved using the AI tool'' may indicate either no change or a negative effect. 
Even where productivity is assessed explicitly, comparability is therefore limited by heterogeneity in how the construct is phrased and elicited.

\begin{mybox}{}
A notable phase contrast is that framework-aligned productivity research primarily recovers human-centered outcomes, whereas explicit productivity assessments focus much more strongly on performance and artifact-related proxies. Human aspects therefore appear substantially underrepresented in explicit productivity assessment.
\end{mybox}

\subsection{\ques{2} Confounders of explicit productivity}

Across the 27 phase-one papers, we identified 44 unique confounding variables in 15 studies.
Of these 44 variables, 21 were reported in a single paper\cite{measuring_the_impact_of_early_2025_ai_on_experienced_open_source_developer_productivity}, which is also one of the few papers in our sample whose primary finding indicates a negative effect of AI tools on developer productivity.
The median number of confounding variables per paper is 1, which stands in sharp contrast to the 44 unique variables identified across the corpus.
The most frequently discussed variables are developer experience (6 papers), AI tool experience (4), project/task complexity (4), project involvement (2), usage approach (2), and usage frequency (2).

This pattern suggests that explicit productivity studies do recognize relevant sources of variation, but not in a sufficiently systematic way to support cumulative interpretation across studies.
Given the small number of papers focusing on different aspects of productivity, we cannot draw generalizable conclusions about how these variables affect specific markers and therefore report only tendencies observed in the included studies.

Focusing on developer experience, the evidence indicates a trend but no clear consensus.
Junior developers appear to benefit more from AI\cite{harnessing_the_potential_of_gen_ai_coding_assistants_in_public_sector_software_development,the_impact_of_ai_on_developer_productivity,ai_assisted_programming_may_decrease_the_productivity_of_experienced_developers_by_increasing_maintenance_burden,measuring_the_impact_of_early_2025_ai_on_experienced_open_source_developer_productivity,generative_ai_and_labour_productivity}, while findings for senior developers are mixed: some studies report smaller gains\cite{harnessing_the_potential_of_gen_ai_coding_assistants_in_public_sector_software_development,generative_ai_and_labour_productivity}, others find no significant effects\cite{measuring_the_impact_of_early_2025_ai_on_experienced_open_source_developer_productivity}, and some suggest either exclusive benefits for experienced developers\cite{significant_productivity_gains_through_programming_with_large_language_models} or declines in their productivity\cite{ai_assisted_programming_may_decrease_the_productivity_of_experienced_developers_by_increasing_maintenance_burden}.
The evidence for AI tool experience and task/project complexity is somewhat more consistent.
Prior experience with the tools is associated with neutral, but mostly positive, impacts\cite{echoes_of_ai,the_space_of_ai,significant_productivity_gains_through_programming_with_large_language_models,using_llm_coding_assistants_to_increase_software_development_productivity_by_33}, while increasing task complexity is associated with less positive productivity impacts\cite{developers_perceptions_on_the_impact_of_chatgpt_in_software_development,harnessing_the_potential_of_gen_ai_coding_assistants_in_public_sector_software_development,the_space_of_ai}.
Thus, even for the most frequently discussed confounders, the overall picture remains mixed.

The remaining 38 confounding variables were each reported only once, although some overlap conceptually with the top six.
For example, programming skills\cite{the_impact_of_ai_on_developer_productivity} and programming language experience~\cite{echoes_of_ai} likely relate to developer experience as sub-factors, while others are distinct, such as prompting behavior\cite{significant_productivity_gains_through_programming_with_large_language_models}.
Many of these variables are also not specific to the settings of their respective studies and overlap with markers derived in \ques{1}.
Examples include AI output accuracy\cite{examining_the_use_and_impact_of_an_ai_code_assistant_on_developer_productivity_and_experience_in_the_enterprise}, AI output speed\cite{examining_the_use_and_impact_of_an_ai_code_assistant_on_developer_productivity_and_experience_in_the_enterprise}, or AI tool limitations\cite{measuring_the_impact_of_early_2025_ai_on_experienced_open_source_developer_productivity}, all of which appear in \autoref{fig:marker-tree}.
This reinforces that the field has not yet stabilized which constructs should be treated as dependent variables, contextual factors, or confounders.

\begin{mybox}{}
Confounding variables are acknowledged, but not standardized. The literature recognizes many potentially important confounders, yet most are discussed only once, which sharply limits cross-study interpretability.
\end{mybox}

Naturally, this analysis does not imply that these 44 variables are the only confounding variables discussed in the papers, as additional variables may have been collected without being explicitly discussed in relation to productivity findings.
Moreover, to avoid blurring the results with unrelated confounders, we excluded the phase-two data from this step and maintain focus on variables discussed alongside explicit productivity findings.

\subsection{\ques{3} Explicit productivity effect directions}

We include the aggregated effect directions (positive, negative, mixed, or neutral/no effect) of phase-one productivity markers and categories as bar charts in the tree representation in \autoref{fig:marker-tree}.
Categories at higher levels aggregate results of lower levels.
We chose this aggregation to maintain readability, as many markers occurred only rarely and separate representations for tendency-based, proportional, and statistically evaluated reports would have been overly sparse.
Results not clearly positive, neutral, or negative were omitted; for example, accepted code size\cite{significant_productivity_gains_through_programming_with_large_language_models} was reported without directional interpretation by the original authors.

Across explicit productivity markers, positive reported effects dominate overall.
However, this aggregated pattern masks a more important result: effect direction varies systematically with reporting format and marker type.
Tendency-based findings are decidedly more positive than statistically evaluated findings, and perceived markers are more positive than objective markers.
The key result is therefore not simply that positive findings outnumber negative ones, but that the apparent positivity of the literature depends, in part, on how effects are measured and reported.

We report the counts of result directions by reporting type (tendencies, proportions, and statistically evaluated effects) and by marker type (perceived and objective) across all markers in \autoref{tab:contingency}.
Positive results were the most common for all three reporting types.
However, the distribution differs substantially across them: tendency-based results show the highest proportion of positive findings, proportional results contain relatively more mixed findings and no negatives, and statistically evaluated effects contain relatively more neutral and negative findings.
Likewise, perceived markers show a higher proportion of positive results than objective markers.
Overall, about two thirds of all results are positive (63\,\%), with 10\,\% negative and 23\,\% neutral.

Given the small sample size, we used pooled Fisher's exact tests to explore associations between result direction and reporting type (tendencies, proportions, and effects).
The omnibus test indicated a significant association ($p < .001$).
Post hoc pairwise Fisher's exact tests with Holm correction showed significant differences between all three reporting types (tendencies vs.\ proportions: adjusted $p < .001$; tendencies vs.\ effects: adjusted $p < .005$; proportions vs.\ effects: adjusted $p < .01$).
We then applied the same pooled exploratory analysis to perceived versus objective markers and again found a significant association ($p < .05$).
By contrast, the association between result direction and gray versus non-gray studies was not statistically significant ($p > .05$), with distributions appearing broadly similar across groups.
Because we extracted multiple results from some papers, these pooled contingency analyses do not fully account for within-paper dependence and should be interpreted as exploratory rather than confirmatory.

\begin{table}[htbp]
    \centering
    \small
    \caption{Result direction by reporting type and operationalization; percentages are row-wise.}
    \label{tab:contingency}
    \begin{tabular}{p{1.2cm}ccccc}
        \toprule
         & positive & negative & neutral & mixed & Total \\
        \midrule\midrule
        Tendency & 44 (81.5\%) & 7 (13.0\%) & 3 (5.6\%) & 0 (0.0\%) & 54 \\
        Proportional & 21 (56.8\%) & 0 (0.0\%) & 11 (29.7\%) & 5 (13.5\%) & 37 \\
        Statistical & 41 (53.9\%) & 10 (13.2\%) & 24 (31.6\%) & 1 (1.3\%) & 76 \\
        \midrule
        Perceived & 64 (71.1\%) & 6 (6.7\%) & 15 (16.7\%) & 5 (5.6\%) & 90 \\
        Objective & 42 (54.5\%) & 11 (14.3\%) & 23 (29.9\%) & 1 (1.3\%) & 77 \\
        \midrule
        Total & 106 (63.5\%) & 17 (10.2\%) & 38 (22.8\%) & 6 (3.6\%) & 167 \\
        \bottomrule
    \end{tabular}
\end{table}

\begin{mybox}{}
The literature is not simply ``mostly positive''. Positivity is strongest in tendency-based and perceived results, whereas statistically evaluated and objective results contain substantially more neutral and negative findings.
\end{mybox}

Put differently, the more interpretively lightweight the reporting format, the more positive the reported productivity picture appears.
This shifts the interpretation of the overall trend.
The central question is not only whether positive findings outnumber negative ones, but whether positivity is robust across stronger and weaker forms of evidence.
In our data, the strongest positive skew appears in tendency-based reports, whereas statistically evaluated effects and objective markers show more neutral and negative results.
The resulting picture is therefore not one of uniformly positive evidence, but of positivity filtered through heterogeneous reporting practices.

Looking at the distributions shown in \autoref{fig:marker-tree}, most marker categories exhibit predominantly positive reported directions.
\emph{Time} is particularly notable, as it is the most frequently reported category and is associated almost exclusively with positive results.
Most time-related markers show positive trends, with the main exceptions being the time until a pull request is merged and the time until a pull request is reviewed or approved.
One study reported both a significant increase in time until merge (classified negative), yet more timely merges (positive)\cite{the_impact_of_generative_ai_on_collaborative_open_source_software_development}.
For time until review/approval, findings were mixed across studies, with one significant increase, one significant decrease, and one non-significant result\cite{the_ai_productivity_paradox,intuition_to_evidence,the_effects_of_generative_ai_on_high_skilled_work}.

Other largely positive categories include \emph{Direct productivity assessment} and \emph{Repository activity}, albeit with slightly more neutral findings than \emph{Time}.
More diverse patterns appear in \emph{Production activity} and \emph{Collaboration activity}, although these categories were reported only rarely (2--3 results each).
\emph{Code quality} is overall more positive than negative, but includes a substantial number of neutral results, with 3 of 16 reports classified as negative.
Negative findings come from two papers reporting more bugs and issues\cite{the_impact_of_generative_ai_on_collaborative_open_source_software_development,the_ai_productivity_paradox}, and one reporting more pull request revisions after the AI tool’s introduction\cite{ai_assisted_programming_may_decrease_the_productivity_of_experienced_developers_by_increasing_maintenance_burden}.

The most negative category is \emph{AI assessment}, with 62\,\% negative reported results (5/8).
These were mostly tendency-based reports concerning AI-tool accuracy, for example open-text responses on hallucinations or incorrect code\cite{the_influence_of_using_llms_on_the_activities_of_a_software_testing_team,significant_productivity_gains_through_programming_with_large_language_models}.
This creates a notable tension in the literature: productivity-related outcomes often trend positive, while evaluations of the AI tools themselves are frequently critical.
Developers may therefore perceive practical gains even while remaining skeptical about tool reliability and correctness.

\begin{mybox}{}
One of the most interesting tensions in the data is that productivity markers often trend positive, while \emph{AI assessment} trends negative. Developers may experience gains in speed or task completion while still judging the tool itself as inaccurate, brittle, or error-prone.
\end{mybox}

Similar to the confounding variable analysis, we restricted this step to explicit productivity markers, because comparability was already substantially constrained by the aggregation of heterogeneous reported effects across studies, marker operationalizations, and reporting types.
Extending the analysis to the broader marker set would have introduced additional conceptual heterogeneity, further weakening the interpretability of the aggregations.

 \section{Discussion and Implications}

\subsection{\ques{1}: Markers of productivity research}
The central contribution of this review is not that research on AI coding assistants and developer productivity is scarce, but that the available evidence is difficult to compare in a cumulative way.
Across the literature, productivity and productivity-related impacts are operationalized through a broad range of markers, while individual studies usually only capture a narrow subset of them.
This indicates that while the field contains numerous relevant findings, there is not yet a coherent evidence base from which more robust conclusions can be drawn easily.

The phase distinction helps interpret what kind of productivity evidence the field currently prioritizes.
Phase one shows a default operationalization of productivity that centers on observable work performance, especially time-, artifact-, and repository-related proxies.
This is understandable, as such markers are often easier to observe, quantify, and compare within individual study designs.
Concurrently, when the broader literature is mapped onto established productivity frameworks such as SPACE, human-centered outcomes, such as satisfaction, psychological aspects, and well-being, occur far more prominently than in phase one.
This suggests that current explicit productivity assessment in software engineering tends to underrepresent the human side of productivity, even though these dimensions are central to productivity effects. 

This matters because software development productivity is not reducible to output and speed alone.
Prior work such as SPACE and DevEx already frames productivity as a multidimensional construct that also includes internal experience, collaboration, and friction in day-to-day work\cite{space,noda_devex_2023}.
Our findings suggest that the corresponding conceptual foundation already exists, though practice remains narrower.
Therefore, the gap is not one of missing theory, but one of shared research practice.
The challenge for the software engineering community is not that every study must measure every productivity dimension at once, but that focused studies should still define and report productivity in ways that make their findings more interpretable relative to one another.

Our taxonomy did not just highlight this gap, it is the first step towards closing it. We grouped markers according to their operationalization rather than by conceptual similarity. This allows researchers to compare two studies not just on the construct, but on how it was measured. They may find whether two studies genuinely disagree or whether their operationalizations were simply incompatible. They may also use the taxonomy to design their studies in a way that makes them inherently better comparable to other related studies with similar markers and goals.

\subsection{\ques{2}: Confounders of explicit productivity}
The findings on confounding variables reinforce the same broader picture.
The literature clearly recognizes that productivity effects depend on context, but it does not yet treat that context in a systematic or comparable way.
Even the most frequently discussed confounders, such as developer experience, AI tool experience, and task complexity, appear in only a limited number of studies, while many others are mentioned only once.
As a result, contextualization exists, but mostly in a fragmented form.

This has two implications.
First, the effects of AI coding assistants on productivity are unlikely to be universal.
They appear to depend on who is using the tool, potentially the task type and the conditions.
The field should therefore move away from asking whether AI coding assistants improve productivity in general, and instead ask under which conditions they help, hinder, or shift work.
Second, the overlap between markers and confounding variables indicates conceptual instability.
Variables such as AI output accuracy or output speed appear in some studies as outcomes and in others as confounding variables.
This makes studies diverge not only in what they measure and how they measure it, but also in how they position the same construct analytically.
For cumulative research, that is a substantial problem.
For the SE community, this suggests that contextual factors should be treated as part of the core study design rather than optional additions.
Not every study needs to cover every possible confounder, but more consistent reporting and justification of key contextual variables would make productivity findings substantially more interpretable across studies.

\subsection{\ques{3}: Explicit productivity effect directions}
At first glance, the literature appears predominantly positive.
Across explicit productivity markers, positive reported directions outnumber neutral and negative ones.
However, our results suggest that this pattern should not be interpreted as a simple answer.
The positivity of the literature is not stable across reporting forms.
It is strongest in tendency-based and perceived findings, whereas statistically evaluated and objective findings contain substantially more neutral and negative results.
Accordingly, the main interpretation is not that the field has converged on a uniformly positive effect, but that the apparent strength of that effect depends in part on how productivity is measured and reported.

This is important for two reasons.
First, some of the most common operationalizations do not allow negative findings to be expressed clearly.
For example, positively framed self-report items may collapse negative outcomes into neutral disagreement.
Second, perceived gains do not necessarily reflect the same construct as observable performance changes.
Future studies should therefore be more explicit about whether they are reporting perceived productivity, observed behavioral change, or statistically tested effects, and should avoid treating these forms of evidence as interchangeable.

A particularly notable tension in the data is that most outcome categories trend positive, while \emph{AI assessment} appears frequently negative. This suggests that developers can experience tangible improvements in speed, task completion, or output volume, yet simultaneously hold a critical view of the tool itself.
For practice, this means that productivity gains and tool quality should not be conflated.
A tool may appear beneficial in narrow output-oriented terms while simultaneously introducing correction effort, uncertainty, or maintenance burden.
For research, it means that evaluations of AI coding assistants should not focus only on speed or throughput, but also consider such downstream effort introduced by imperfect suggestions.
Taken together, the findings from \ques{3} support a more careful interpretation of the current evidence base.
The literature points to recurring positive patterns, but not to context- or method-independent conclusions about productivity effects.

\section{Threats to Validity}

Our synthesis involves interpretation-dependent coding decisions. In particular, normalizing markers and confounders required semantic judgment, which may have introduced labeling inconsistencies despite calibration and discussion among authors. We also captured confounding variables only when studies explicitly related them to productivity findings. Therefore, collected but not discussed confounders are absent from the synthesis, but this also reflects inconsistent reporting in the field. In addition, separating explicit productivity studies from other outcomes within the same paper imposed an analytic distinction not always made by original authors, although we mitigated this by revisiting phase-one papers during phase two. 

Because the corpus includes peer-reviewed papers, preprints, and industry reports, some aggregated patterns may also reflect variation in rigor and reporting quality across source types. As with any SLR, relevant studies may have been missed despite our multi-source search strategy and control-set validation. Further, applying SPACE in phase two required interpretive judgment in boundary cases. 

Our search combined venue-based screening of 3,746 papers across twelve major SE venues, Semantic Scholar API queries, and manual Google Scholar searches, but did not include snowballing or further database coverage, which could increase the paper pool further. 
However, any resulting gaps in collected papers are unlikely to alter the central finding, given the pronounced heterogeneity and fragmentation of productivity research already evident in the corpus.
Finally, many individual markers were reported only in a small number of studies, limiting the strength of directional summaries. Thus, we aggregate mainly at category level and treat statistical analyses as exploratory rather than confirmatory.
\section{Conclusion}

This literature review synthesized how software developer productivity with AI coding assistants is currently studied.
Across 45 papers and reports, we identified a broad but fragmented evidence base with many different markers, narrow study-level operationalizations, and limited cross-study comparability.
Explicit productivity research focuses primarily on output-, time-, and artifact-oriented proxies, whereas the broader framework-aligned literature makes human-centered dimensions such as satisfaction, psychological aspects, and well-being much more visible.
This suggests that current productivity assessment in software engineering captures important parts of the construct, but not yet the full picture.

Our findings further show that reported productivity effects are often positive, but that this positivity depends on how evidence is measured and reported. Perceived and tendency-based results are more positive than objective and statistically evaluated findings, while contextual factors such as developer experience, tool experience, and task complexity are recognized only inconsistently. 

Taken together, the current literature does not point to a single stable answer to whether AI coding assistants improve productivity in general. Instead, it points to a field that already contains many valuable findings, but still lacks a sufficiently shared, multidimensional, and context-sensitive research practice. 
The next key step is therefore not only to produce more productivity studies, but to make them more comparable through clearer specifications of which productivity dimensions are assessed, more consistent reporting and assessing of markers, broader inclusion of human-centered outcomes, and more systematic analysis of confounding variables. 
If developer productivity with AI coding assistants is a puzzle, the field has already collected many important pieces.
This study adds a precise account of which 
pieces are missing, how they look, and where they fit.
The next challenge is to make them fit together.

\subsection*{Data Availability Statement}
All data, analysis scripts and additional visualizations can be found at our supplementary material published via Zenodo\cite{wittig_2026_21352349}; URL: \url{https://doi.org/10.5281/zenodo.19219046}

\subsection*{Acknowledgements} The authors acknowledge the financial support by the Federal Ministry of Research, Technology and Space of Germany and by Sächsische Staatsministerium für Wissenschaft, Kultur und Tourismus in the programme Center of Excellence for AI-research ``Center for Scalable Data Analytics and Artificial Intelligence Dresden/Leipzig'' (project identification number: ScaDS.AI) and by the European Social Fund (ESF) together with the German state of Saxony under grant number A.100760692 (project: Teaching-AI).

\bibliographystyle{ACM-Reference-Format}
\bibliography{review,related}

\end{document}